# Atomic scale imaging of the negative charge induced by a single vanadium dopant atom in monolayer $WSe_2$ using 4D-STEM

*Djordje Dosenovic[1], Kshipra Sharma[1], Samuel Dechamps[1,2], Jean-Luc Rouviere[1], Yiran Lu[3], Antoine Mordant[1], Martien den Hertog[3], Luigi Genovese[1], Simon M.-M. Dubois[2], Jean-Christophe Charlier[2], Matthieu Jamet[4], Alain Marty[4], Hanako Okuno[1]\**

[1]Univ. Grenoble Alpes, CEA, IRIG-MEM, 38000 Grenoble, France

[2]Institute of Condensed Matter and Nanosciences, Université catholique de Louvain (UCLouvain), 1348 Louvain-la-Neuve, Belgium

[3]Univ. Grenoble Alpes, CNRS-Institut Néel, F-38000, Grenoble, France

[4]Univ. Grenoble Alpes, CEA, CNRS, Grenoble INP, IRIG-SPINTEC, 38000 Grenoble, France

# ABSTRACT

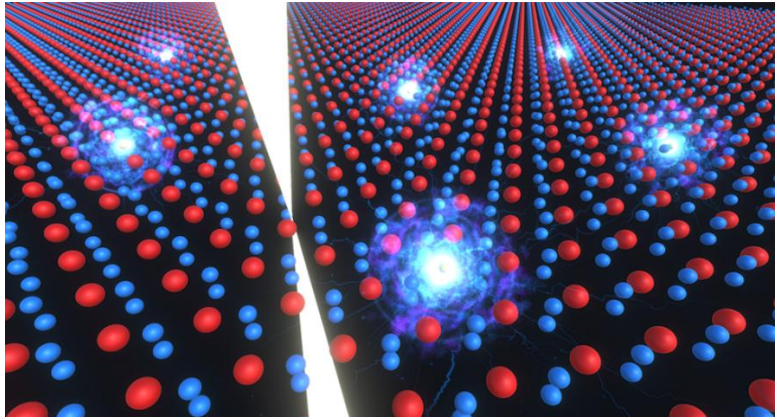

There has been extensive activity exploring the doping of semiconducting two-dimensional (2D) transition metal dichalcogenides in order to tune their electronic and magnetic properties. The outcome of doping depends on various factors, including the intrinsic properties of the host material, the nature of the dopants used, their spatial distribution as well as their interactions with other types of defects. A thorough atomic-level analysis is essential to fully understand these mechanisms. In this work, vanadium doped $WSe_2$ monolayer grown by molecular beam epitaxy is investigated using four-dimensional scanning transmission electron microscopy (4D-STEM). Through center of mass-based reconstruction, atomic scale maps are produced, allowing the visualization of both the electric field and the electrostatic potential around individual V atoms. To provide quantitative insights, these results are successfully compared with multislice image simulations based on ab initio calculations, accounting for lens aberrations. Finally, a negative charge around the V dopants is detected as a drop in the electrostatic potential, unambiguously demonstrating that 4D-STEM can be used to detect and to accurately analyze single dopant charge states in semiconducting 2D materials.

# INTRODUCTION

Owing to their versatile electronic properties and reduced dimensionality, layered two-dimensional (2D) materials have emerged as promising candidates for exceptionally compact devices[1]. Among these materials, transition metal dichalcogenides (TMDs) have received special attention due to their often semiconducting nature and the ability to fine-tune their electronic structure, making them ideal for innovative logic and memory devices[2–4]. In recent years, there has been significant progress in the synthesis of 2D materials, enabling the creation of materials tailored to specific properties[5]. However, the properties of these synthesized materials are significantly influenced by intrinsic and extrinsic structural imperfections, such as vacancies, grain boundaries, and dopants. The resulting functionality of the corresponding 2D material is intricately linked to the electronic properties of these defects, as well as their concentration and distribution within the sample. Molecular beam epitaxy (MBE) has emerged as a valuable process for enhancing the complexity of scalable engineered 2D materials. Notably, it enables precise control over the introduction of dopant atoms into 2D layers, thus offering a means to finely adjust the material properties[6,7]. In the context of otherwise non-magnetic TMDs, doping with magnetic elements like V, Mn, Fe, and Co in low concentrations has the potential to induce long-range ferromagnetic ordering. This leads to the creation of what are known as diluted 2D magnetic semiconductors, where the ferromagnetic ordering can be controlled by a gate voltage[8–10]. The theoretical predictions of long-range ferromagnetic ordering in such systems[11] have been experimentally confirmed[12]. However, it is worth noting that the magnetic moment induced by dopant atoms is expected to be suppressed when the dopants carry a negative charge due to the occupation of defect-induced states[13,14]. Therefore, the ability to investigate the atomic-scale structure, the local chemical environment, and the charge state of the dopant is crucial for a comprehensive understanding of the properties of the 2D materials. While local probe techniques, namely scanning tunneling microscopy/spectroscopy (STM/STS), are commonly used for the analysis of local electronic structure, the information coming from electronic and atomic structures is difficult to dissociate. The chemical identification of impurities and other defects using such methods is often challenging and ambiguous[15]. In this context, aberration-corrected scanning transmission electron microscopy (AC-STEM) equipped with analytical spectroscopy techniques such as electron energy loss spectroscopy (EELS) stands out as a most effective tool for structural and chemical analysis at the atomic scale.

Today, with the development of the differential phase contrast (DPC) imaging technique additional information going beyond the identification of chemical species and their positions can be accessed. The DPC technique measures the deviation of the electron beam off the optical axis induced by the in-plane components of local electric and magnetic fields[16]. This technique was first demonstrated using a segmented detector, allowing for an approximate estimation of the electron beam deflection. Recent developments in pixelated electron detectors enable increased measurement capabilities. In the so-called four-dimensional scanning transmission electron microscopy (4D-STEM) acquisition mode[17], a 2D diffraction pattern is collected for each beam position allowing a more precise measurement of the beam deflection. The deviation of the center of mass (CoM) of the transmitted electron beam in each collected diffraction pattern can be directly related to the local electric field[18,19]. The CoM approach is therefore rather straightforward, contrary to other phase retrieval techniques such as ptychography that requires more sophisticated computational processing[20]. By solving the Poisson's equation, other associated fundamental quantities such as the electrostatic potential and charge density can be obtained from the electric field[19]. Nevertheless, the DPC technique is not yet used in the characterization of 2D materials on a routine basis. Previous demonstrations have been limited to model systems, i.e. pristine and point defect structures in exfoliated $MoS_2$[21] and line defects in $MoS_2$ and $WS_2$ created by electron irradiation inside the microscope[22,23]. The imaging of the electric field of single Si dopants in monolayer graphene grown by chemical vapor deposition (CVD) has also been used to distinguish between two different dopant coordination states[24]. The widespread application of the method in studying real synthesized materials is hindered by experimental difficulties related to long acquisition times, sample stability under the beam and transfer-related polymer contamination, as well as challenges in quantitative interpretation of the reconstructed electric field, potential and charge density images. Importantly, the influence of residual lens aberrations on the electric field images reconstructed by DPC-CoM has often been mentioned but rarely discussed in detail[25,24,22]. Recently, the probe size and aberration effects have been studied to quantitatively resolve the electron charge density of individual atomic columns or single atoms in TMDs, pointing out the current limitation of the technique to access fundamental material information such as the valence electron distribution[26].

In this work, we apply the 4D-STEM based CoM reconstruction to analyze the long-range modulation of the electrostatic potential observed within a V-doped $WSe_2$ monolayer. We start by establishing a workflow for quantitatively analyzing 4D datasets. This is achieved by incorporating density functional theory (DFT)-based STEM multislice simulations, which take into account the residual aberrations inherent in the experimental data. To estimate these residual aberrations, we conduct a ptychographic probe reconstruction for each of the experimental 4D datasets. We apply this methodology to achieve atomic-resolution mapping of the electric field within a V-doped WSe$_2$ monolayer, grown directly on graphene/Pt using molecular beam epitaxy (MBE). Finally, the local electric field and the potential around single atom V dopants are measured using 4D-STEM CoM reconstruction, where a potential drop induced by the negatively charged dopant atoms is detected, in analogy to previous STM observations on a similar system[27].

## RESULTS AND DISCUSSION

The V-doped WSe$_2$ was grown by MBE on CVD graphene monolayer as grown on Pt/Si substrate, using the same growth process shown in a previous work to synthesize the V-doped WSe$_2$ on graphene/SiC for the STM study[27]. The WSe$_2$/graphene hetero-stack was transferred by polymer-assisted wet transfer using the electrochemical method[28], followed by surface cleaning. The STEM imaging, EELS spectrum imaging and 4D-STEM acquisition are performed in low-voltage (80 kV) condition to minimize the creation of chalcogen vacancies in TMDs[29]. Experimental 4D-diffraction datasets for the CoM analysis are registered using a fast pixelated camera. Simulated diffraction patterns are obtained using STEM multislice simulations where the electrostatic potential of the (V-doped) WSe$_2$ monolayer estimated by DFT is used to model the specimen. The electron beam is simulated taking into account microscope conditions such as acceleration voltage, convergence angle and lens aberration coefficients. The center of mass of the transmitted electron beam is measured from both the experimental and simulated diffraction patterns, and subsequently converted into the projected electric field using Ehrenfest theorem[18,19]. Finally, the experimental and simulated images of projected electrostatic potential are generated by integration in Fourier space[30]. **Figure 1a** shows the high angle annular dark field (HAADF) image of V-doped WSe$_2$ on graphene that exhibits relatively clean and highly crystalline structures, where two typical structural anomalies are found; individual point defects at the W sites and triangular inversion domains (TIDs). The HAADF image confirmed the presence of a light atom replacing W atom at each point defect (**Figure S1**), suggesting successful vanadium doping as demonstrated elsewhere[27,31]. TIDs are rarely observed in pristine WSe$_2$ due to their high formation energy[32]. The addition of vanadium to the growth system has been demonstrated to lower the formation energy[33] and promote the appearance of TIDs consisting of stable 4|4P line defects, so-called mirror twin boundaries (MTBs) under the Se deficient condition, which is often unavoidable in MBE growth[34]. **Figure 1b** illustrates the atomistic model of MTBs observed in V-doped WSe$_2$, as relaxed by DFT structural optimization. **Figure 1c** and **d** show 2D images of the magnitude of the projected electric field and the electrostatic potential around a MTB reconstructed by the CoM method.

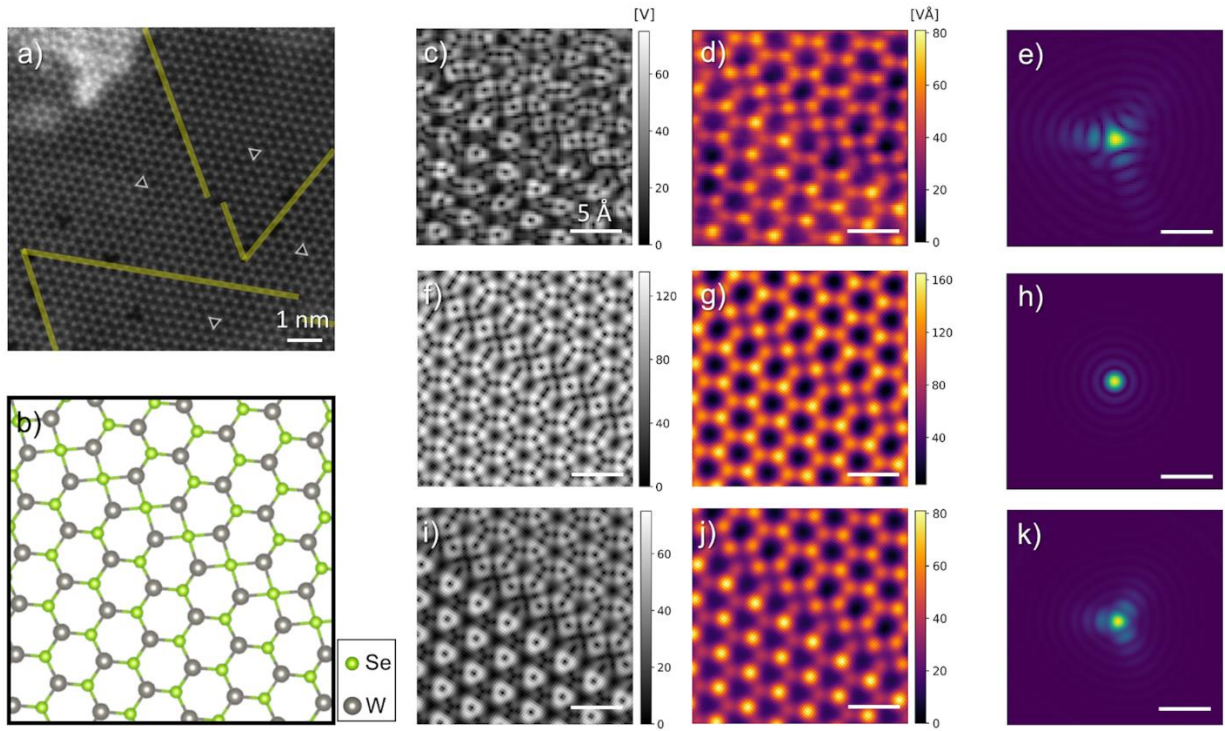

*Figure 1. (a) HAADF image of V-doped WSe$_2$ sample. MTBs are indicated by yellow lines. (b) DFT-relaxed model of MTB in WSe$_2$ used for the image simulations. (c) Experimental map of projected electric field around a MTB reconstructed by the CoM method and (d) the corresponding projected potential map. (e) Ptychographic reconstruction of the electron probe used during the acquisition of the 4D dataset. (f) Projected electric field map reconstructed from a simulated 4D dataset without any lens aberrations and (g) the corresponding projected potential map. (h) Simulated aberration-free electron probe. (i) Projected electric field map and (j) potential map reconstructed from simulated 4D dataset with defocus = -9 nm and A2 = 220 nm. (k) Simulated aberrated electron probe. All the scale bars in (c-k) correspond to 5 Å.*

Given the symmetry of the 4|4P domain junction, it is expected that the electric field and the electrostatic potential of $WSe_2$ will exhibit identical contrast patterns on the two sides of the MTB with a 60° rotated mirror crystal symmetry. However, the reconstructed images of the projected electric field and potential present substantially different contrast. Since the grain boundary structure is known from the HAADF imaging and the energy stability arguments[32], the positions of W and 2Se in the reconstructed images are unambiguously determined (**Figure S2**). The intensities of W atoms in the upper right corner in **Figure 1d** are higher than the intensities of neighboring 2Se columns, while the intensity ratio of these two columns is inversed on the opposite side of a MTB. This variation in intensity as a function of crystal orientation has often been studied in atomic resolution HAADF images, as a consequence of residual aberrations[35–37]. **Figure 1f,g** depict the magnitude of the projected electric field and the electrostatic potential reconstructed from the DFT-based STEM multislice simulation, including the main microscopy parameters but in the aberration-free condition (**Figure 1h**).

Indeed, without probe aberrations the reconstructed projected electric field and potential exhibit a mirror symmetry around the MTB as expected. In order to survey the residual aberration present in the probe, we perform the ptychographic probe reconstruction directly from the experimental 4D dataset used to obtain the electric field and potential shown in **Figure 1c,d**. The reconstructed probe intensity (**Figure 1e**) exhibits a triangular shape commonly attributed to the effect of residual three-fold aberrations (i.e. A2 and D4)[35,38]. Furthermore, the overall magnitudes of electric field and potential predicted by the aberration-free simulation exceed the experimental values by about a factor of 2, even after taking into account the spatial incoherence of the probe, as previously observed[22,24]. This attenuation in intensity is assumed to be due to defocus and other spherical aberrations that contribute to the beam broadening[25].

**Figure 2a** shows simulations of the atomic column intensity dependence on defocus. The defocus indeed attenuates the intensities of both the W and 2Se atomic columns while their intensity ratio remains constant. The effect of three-fold aberrations depends on their magnitude as well as on the azimuthal angle φ, which can be understood as the orientation of the triangular probe. A series of multislice simulations are performed with different orientations of A2 astigmatism while the orientation of the crystal is kept constant (**Figure 2b,c**). The azimuthal angle of the aberration is varied from 0° to 120° with an increment of 10° and the average potential around the 2Se and W atomic columns is evaluated for each simulated reconstructed projected potential. The intensities of the 2Se and W columns shown in **Figure 2b,c** are strongly dependent on the azimuthal angle $\varphi_{A2}$. In a certain range of orientations ($10° < \varphi_{A2} < 50°$) the W atoms appear brighter than 2Se atomic columns, leading to the contrast inversion, whereas HAADF imaging under the same beam conditions is found to be more resistant to the effect of A2 astigmatism (**Figure S3**).

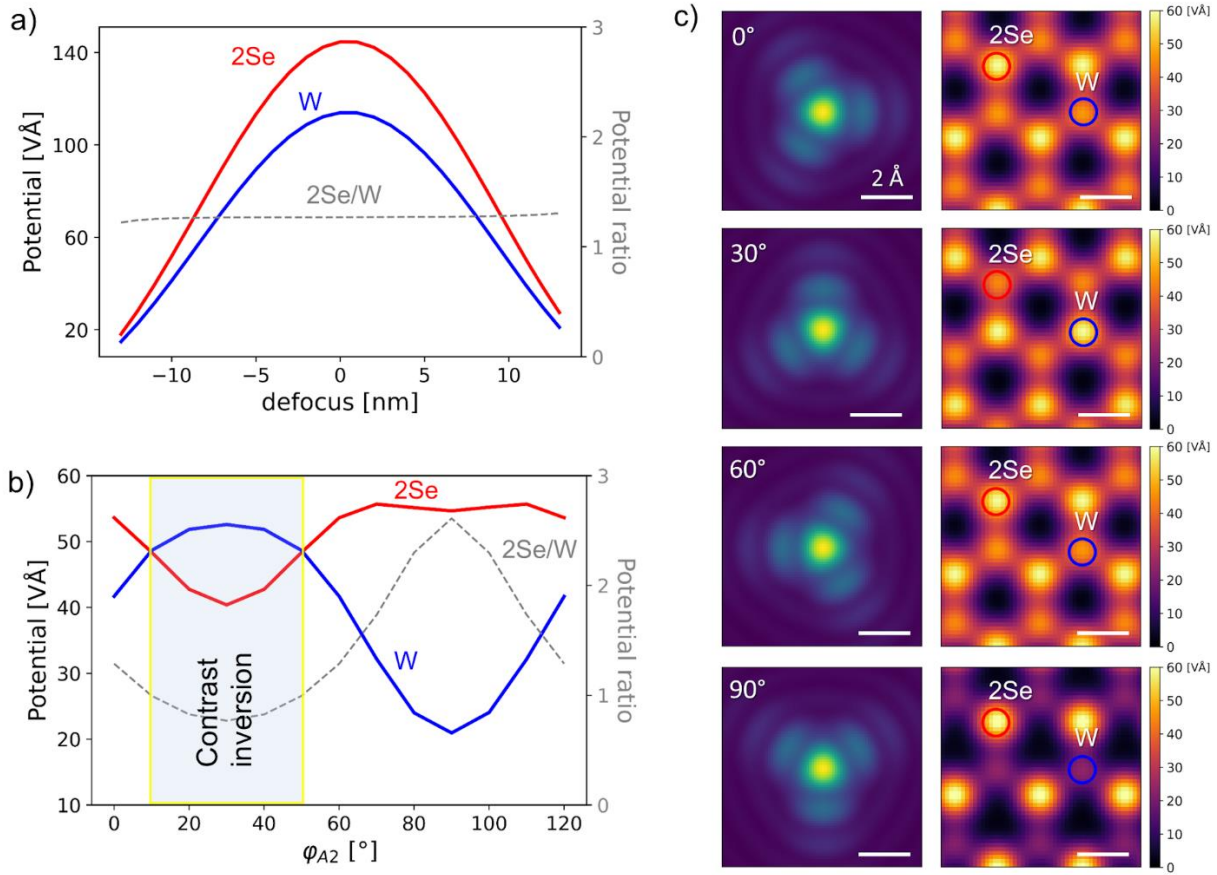

*Figure 2.* *(a) Average projected potential around 2Se (red) and W (blue) atomic columns and potential ratio (dashed gray) from a series of multislice simulations with different values of defocus and without any additional aberrations. The reference zero defocus is defined on the W plane and the positive focus is defined as underfocus. (b) Average projected potential around atomic columns and potential ratio from a series of multislice simulations with the fixed value of defocus (def = -9 nm) and the magnitude of threefold astigmatism (A2 = 220 nm) and the varying azimuthal angle of A2 aberration. (c) The typical examples of simulated probe and the reconstructed projected potential for the characteristic values of A2 azimuthal angles: $\varphi_{A2}$=0°, 30°, 60°, 90°. All the scale bars in probe and potential images correspond to 2 Å.*

In practice, the orientation of the electron probe is quite stable during an experiment. However, the 2D layers are often polycrystalline, and the orientation of the crystal with respect to the electron probe can be arbitrary. For this reason, the contrast will strongly depend on the local orientation of the crystal at the region of interest. As already shown in **Figure 1**, an acquisition around a 4|4P MTB results in a significantly different contrast at the two sides of the domain boundary since the two crystallites are misoriented by 60° and the reconstructed projected potential on the two sides of the MTB would correspond roughly to the potential reconstructed from the simulations with $\varphi_{A2} = 30°$ and $\varphi_{A2} = 90°$, shown in **Figure 2c**.

In order to include the experimentally used microscope parameters in the simulations, the angle of the apparent three-fold aberration is measured directly from the reconstructed probe image shown in **Figure 1e**. The virtual defocus and magnitude of A2 are then fine-tuned by comparing the simulated intensities of the W and 2Se columns with those obtained experimentally in defect-free reference regions. **Figure 1i,j** show the images of the projected electric field and electrostatic potential, respectively, reconstructed from the simulation taking into account these two parameters (defocus = -9 nm and A2 = 220 nm), giving the electron probe intensity shown in **Figure 1k**. The overall intensities are now in agreement with the experiment due to defocus effects, while the contrast inversion between the W and 2Se atomic columns on the two sides of the MTB is achieved by setting corresponding A2 astigmatism. Here, it should be noted that the three-fold aberration is considered only with A2 astigmatism as single parameter. The obtained value (A2 = 220 nm) might include other higher order aberrations such as D4, which may explain the value much higher than A2 usually measured in the microscope. Similarly, the fixed defocus might include other spherically symmetric aberrations (focal spread, 3rd order (C3) and 5th order (C5) spherical aberrations). The combined effect of all aberrations is thus approximated by fitting only two virtual coefficients (defocus and A2). This approach significantly reduces the number of variables in the simulations without a major loss on the quantitative comparison and allows to acquire a reliable simulation for each experimental dataset.

This quantitative analytical process is applied to study the local potential state induced by single V dopant atoms incorporated into $WSe_2$ monolayer deposited on graphene monolayer. **Figure 3a** shows the HAADF image of the V-doped $WSe_2$ layer, where large concentration of point defects at W site is observed both in pristine regions and in mirror domain boundaries. Additional local chemical analysis by EELS spectroscopy unambiguously confirms the presence of V single atom dopants substituting W atoms (**Figure 3b**). A 4D-STEM acquisition is then performed around a single V atom, from which we reconstruct the real space images of the projected electric field and projected potential (**Figure 3d,f**). The V dopant exhibits higher visibility in the reconstructed electric field and electrostatic potential maps than in the HAADF image (**Figure 3a**) and the ADF image virtually reconstructed from the same dataset (**Figure 3c**). The use of DPC-CoM for the simultaneous visualization of light and heavy elements has already been demonstrated and discussed[39]. According to our results presented in **Figure 2**, the intensity ratio 2Se/W as well as the visibility of light elements in the potential map varies as a function of the residual three-fold astigmatism and defocus.

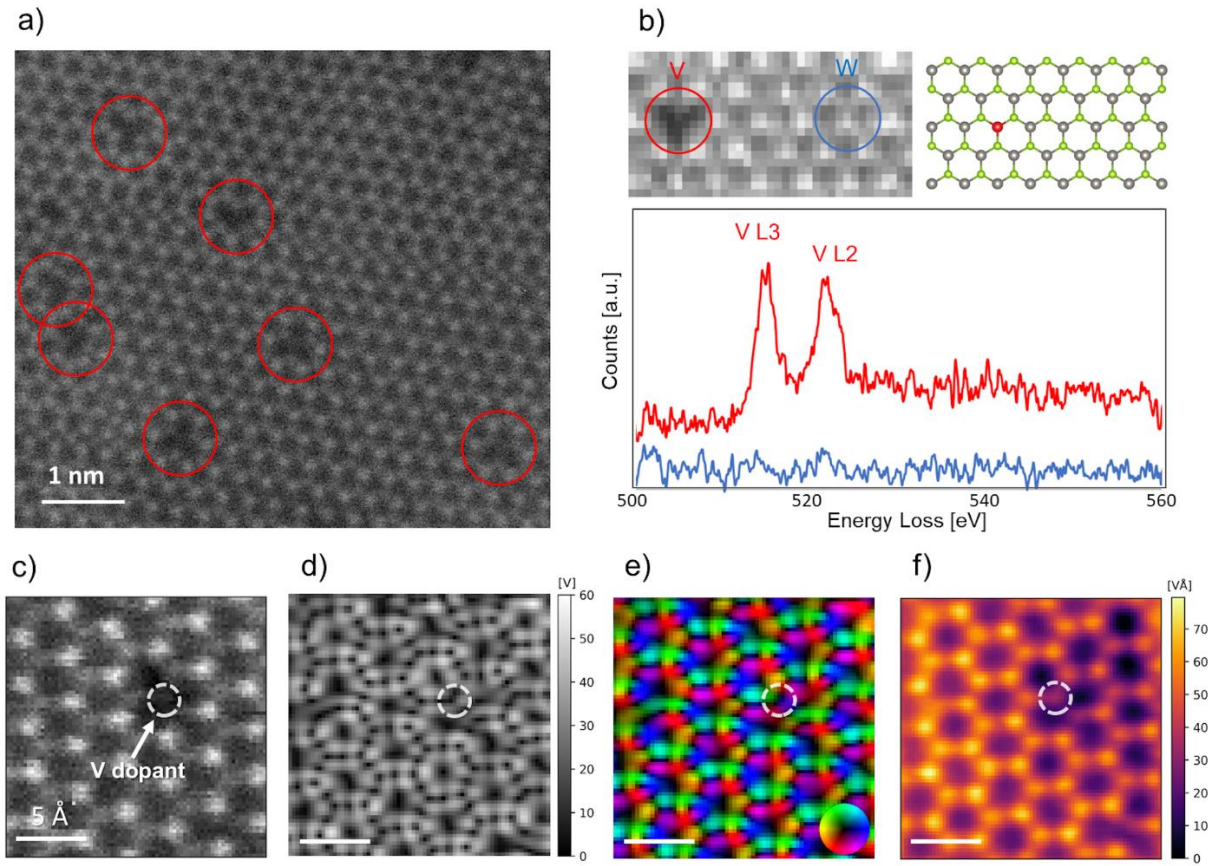

*Figure 3. (a) HAADF image showing the distribution of point defects (red circles) in V-doped WSe$_2$. (b) EELS spectrum around a point defect at tungsten site (red) indicating incorporation of vanadium dopant and the reference spectrum around tungsten atom (blue). (c) Virtual ADF image reconstructed from a 4D dataset acquired around vanadium dopant. (d) Magnitude of projected electric field, (e) its vector whose direction is given by the color wheel and (f) the corresponding projected electrostatic potential reconstructed from the 4D dataset.*

Here, the potential map is affected by the residual three-fold astigmatism effect and the W site gives more intensity than the 2Se site, which is confirmed by the atomic positions attributed in the corresponding virtual ADF image.

A peculiar potential drop around dopant atoms is detected in the reconstructed potential image, **Figure 3f**. There have been several works reporting that substitutional V atoms in WSe$_2$ induce states located close to the valence band maximum, whose net negative charge has been measured by local probe techniques[27,31,40]. The potential drop observed here is supposed to be a consequence of this charge state. To investigate this assumption, an experimentally obtained projected electrostatic potential map (**Figure 4a**) is compared with those reconstructed from two simulated datasets (**Figure 4b,c**). While the first multislice simulation estimates the potential using the independent atom model (IAM), with a neutral dopant, the second model is based on density functional theory (DFT), where a net negative charge is assigned to vanadium.

Our 4D-STEM measurement gives access to three projected physical quantities, namely the electric field, the potential and the charge density. Among the three, we focus on the electrostatic potential for the comparison with the DFT-based simulations. Stemming from its vectorial nature, the 2D image representation of the electric field as measured directly by the CoM method is not straightforward (**Figure 3d,e**). On the contrary, both the electrostatic potential and the charge density are scalar fields obtained by integration and divergence of the electric field, respectively, which can be easily represented in a real-valued 2D image. Additionally, charge effects appear more distinctly in the potential map than in the corresponding charge density map (**Figure S4**). Finally, integration has a smoothing effect that reduces numerical noise, while divergence has the opposite effect. All these arguments go in favor of using the images of the reconstructed projected potential for the analysis of the dopant charge states. We first identify the values of defocus and A2 astigmatism from the dataset used in **Figure 4**, which reproduce the experimental conditions, as explained above. In contrast to the dataset shown in **Figure 3**, the analyzed area exhibits a different aberration effect, where the intensity of the W columns is lower than that of the 2Se columns. Subsequently, we compare the reconstructed potential image computed from the IAM and DFT with the experimental parameters including the aberrations estimated from the experimental data. While the former simulation predicts the intensities of the individual W, 2Se and V atoms with satisfactory accuracy (**Figure 4e**), it fails to reproduce the experimentally observed background intensity variation around the V atom (**Figure 4d**). On the other hand, localizing a negative charge of $q = 0.45e$ on the V dopant in the DFT-based 4D-STEM multislice simulation results in a reconstructed potential that exhibits the background intensity gradient as observed in the experimental data (**Figure 4f**).

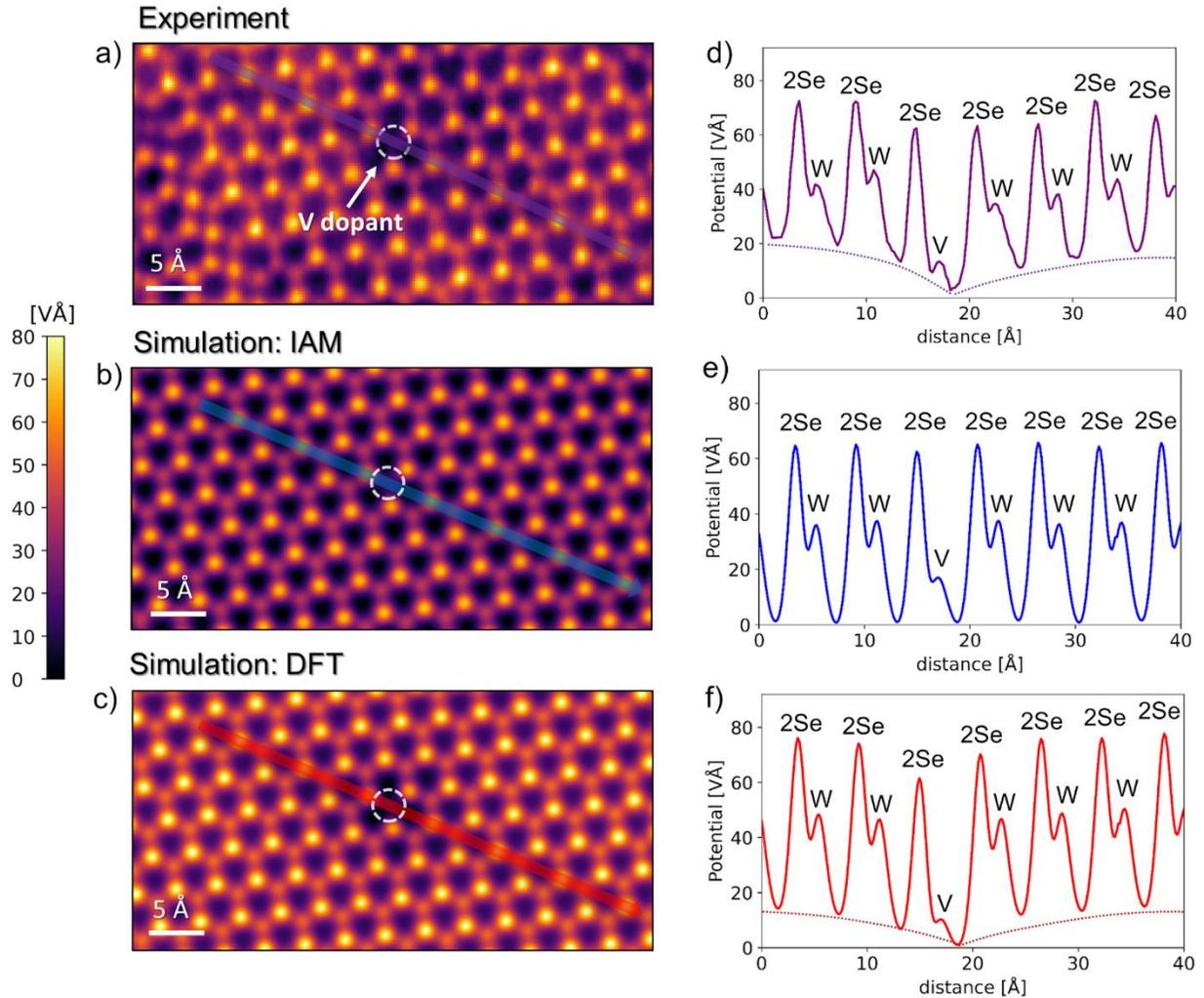

*Figure 4. (a) Experimental image of the projected electrostatic potential around a V dopant atom. (b-c) Images of the projected potential reconstructed from multislice simulation with def = -9 nm, A2 = 120 nm where the input potential of V doped WSe$_2$ is generated from IAM (neutral V dopant) and DFT (charged V dopant), respectively. (d-f) Line profiles through the regions indicated in the potential images (a), (b) and (c), respectively. The dashed lines point out the variation of the background potential around a V dopant atom.*

Therefore, we attribute the experimentally observed potential drop to the presence of negative charge around the V dopant atom. Similar doping effects have been detected by STM in WSe$_2$/graphene/SiC systems[27,31,40], where the charge transfer originates from the heavily doped graphene on SiC[27]. Since the observed WSe$_2$/graphene stack is a freestanding heterostructure that was transferred from its growth substrate, it is not clear whether such a charge transfer mechanism would persist after wet transfer from the native substrate. Alternatively, the charging of V dopant atoms could also occur during the illumination by the electron beam[41,42]. Although the negative charge effects are generally observed around most of the single V dopants in our sample, some V atoms are found not to induce this effect. We also observed an impurity atom at a W site, indicated in **Figure S5**, whose intensity in the potential image corresponds to the intensity of a V atom, although no potential drop is reported in its vicinity. In consequence, this point defect is identified as a neutral V dopant. The charge discrepancy between the two impurities might be explained by the different local electrostatic environments[43]. The presence of a MTB, known to induce local 1D metallic feature in semiconducting TMDs[44], near the neutral dopant may act as a charge trap, while the combination of low screening and long-range Coulomb interaction would lead to complex doping effects between chalcogen vacancies and other V dopants. The doping effect of V single atoms including their geometrical environment should be further explored using the CoM-based quantitative analytical process demonstrated in this work.

## CONCLUSIONS

To summarize, this work delves into a quantitative analytical process based on the CoM method using 4D-STEM and applies it to map the electrostatic potential around vanadium dopants incorporated in WSe$_2$. We have identified a significant impact of residual spherically symmetric aberrations and three-fold aberrations under standard acquisition conditions. These aberrations have the potential to cause misinterpretation of CoM reconstructed images. Consequently, it highlights the importance of comparing experimental results with simulations that consider the dominant residual aberration terms and other microscope parameters. Indeed, when the experimentally obtained CoM reconstructions are compared to those generated from DFT-based multi-slice simulations, mimicking the experimental beam conditions, a quantitative agreement is found. The electrostatic potential mapping around individual dopant atoms reveals a distinct potential drop associated with certain dopant species. This potential drop is successfully reproduced in DFT-based simulations, indicating the localization of a net negative charge at the V dopant atom, thereby demonstrating single charge detection sensitivity using the 4D-STEM CoM method. This finding opens up exciting possibilities for utilizing the CoM technique in screening defect charge states and studying single charge related phenomena. Furthermore, the insights gained from this research provide valuable input for theoretical calculations and future studies aimed at elucidating why the material exhibits the observed charge state and what implications it holds for resulting material properties and magnetic ordering.

**METHODS**

**Sample growth.** The film was grown by molecular beam epitaxy under ultrahigh vacuum (UHV) in a reactor with a base pressure of $5 \times 10^{-10}$ mbar. Prior to the growth, the Gr/Pt substrate was annealed under UHV at 800°C during 30 minutes. The substrate temperature was kept at 340°C during the film growth and W, V were co-evaporated using electron guns at 0.025 Å/s and 0.0025 Å/s deposition rates respectively as monitored by quartz balances. Se was evaporated from a Knudsen cell at a pressure of $1 \times 10^{-6}$ mbar measured at the sample position thanks to a retractable flux gauge. We deposited the equivalent of 0.9 ML of $WSe_2$ and 0.2 ML of $VSe_2$ corresponding to a V concentration of 10 %. After the growth, the sample was annealed at 800°C during 15 minutes under Se flux. After cooling down to room temperature, a 10 nm-thick amorphous selenium layer was deposited.

**Scanning Transmission Electron Microscopy.** Prior to the observation in the microscope, the as-grown sample is spin coated with polystyrene layer at 3000 rpm. The coated stack of $WSe_2$/graphene is then detached from the native Pt substrate by the electrochemical wet transfer method. The floating sample is fished on TEM grid (C-flat$^{TM}$) with 2µm holes and the polystyrene layer is dissolved in toluene. The sample is further rinsed in acetone and isopropanol and left to dry at room temperature. Subsequently, the Se cap is removed by sublimation during heating in vacuum at 250°C for 1h using Gatan heating station and holder. The sublimation of the Se cap additionally contributes to the removal of polymer residues from the sample surface. The sample is observed using aberration corrected FEI Titan Ultimate operating at 80 kV to minimize the beam induced knock-on damage. A convergence semi-angle of 20 mrad was used for all the STEM imaging. The acquisition of HAADF images is done using a beam current of 30 pA and dwell time of 2 µs while the detector captures electrons scattered above 55 mrad. For EELS analysis, we used a GIF Quantum spectrometer (Gatan) operating at 80 kV. The EELS collection semi-angle is 50 mrad. Typical electron beam current is 30 pA. The exposure time is 0.1 s/pixel. The 4D-STEM data is acquired using a dwell time of 0.8 ms and a beam current of 16 pA. The diffraction patterns are collected using Medipix3-based Merlin camera (256x256 pixels) in continuous 12-bit single pixel mode at a $t_0$ threshold of 30 keV.

**Data treatment and 4D-STEM simulations.** The center of mass of each diffraction pattern in the acquired 4D dataset is determined using py4dstem[30] open-source python package and the projected electric field is calculated as detailed in ref[19]. The projected potential is then calculated by the integration in Fourier space and the zero padding factor of 4 was used to assure the periodic boundary conditions and to prevent edge artifacts. The electric field images shown in the manuscript are filtered in Fourier space using a low-pass filter mask with a radius of 38 mrad in order to remove the noise and facilitate the comparison with the electric field images reconstructed from noise-free simulated data.

The multislice simulations are performed using abTEM simulation package[45]. The electrostatic potential that serves as input in the simulations is obtained by GPAW[46] as explained below. The independent atom model input potential is obtained using Lobato parametrization[47]. The electron beam in the simulations is defined by setting the acceleration voltage of 80 kV and the semi-convergence angle of 20 mrad corresponding to the conditions used for the experimental acquisitions. Additionally, the probe aberrations are added, namely defocus and three-fold astigmatism. The input potential for the multislice simulations is cut into slices with a thickness $d_z$ = 0.2 Å. The simulated 4D dataset is then treated in the same way as the experimental datasets as previously explained and the obtained reconstructed maps are convolved with a Gaussian of FWHM = 1 Å to account for the spatial incoherence[48].

**Ptychography reconstruction.** Ptychography reconstruction was performed using a home-made python script reproducing the MATLAB code used in Ref[49]. This code used the ePIE method[50] to reconstruct the complex incident probe.

**Density functional theory.** First principles calculations are performed based on density-functional theory (DFT), norm-conserving pseudopotentials and pseudo-atomic localized basis functions, using the GPAW software package[46]. The Perdew-Burke-Ernzerhof[51] implementation of the generalized-gradient approximation functional is used to describe the exchange-correlation interaction. Applying the Monkhorst-Pack scheme[52], the electronic structure of TMDs are computed with a k-points sampling equivalent to a 12 × 12 × 1 mesh in the unit cell. To avoid spurious interaction, an interlayer distance of 25 Å is employed between the periodically repeated images. The atoms are characterized using a basis set defined from a linear combination of atomic orbitals with two radial functions per valence state and a polarization function, namely the double-zeta polarized. The computations yield 3.34 Å for the lattice parameters of $WSe_2$, in good agreement with the literature[53]. Similarly, the estimated band gap of 1.53 eV is in agreement with DFT calculations[54]. The discrepancy with the experimentally reported values of 2.08 eV[55] originates from DFT notoriously underestimating the band gaps in crystals[56]. In order to model negatively charged V dopants, an attractive point charge potential is located at the defect site. This method allows to simulate charge localization by avoiding standard delocalization errors related to semi-localized functionals in DFT[57].

ACKNOWLEDGEMENTS

We acknowledge the French National Research Agency through the MAGICVALLEY project (ANR-18-CE24-0007). The LANEF framework (No. ANR-10-LABX-0051) is acknowledged for its support with mutualized infrastructure. This project received funding from the European Research Council under the European Union's H2020 Research and Innovation programme via the e-See project (Grant No. 758385). S. D., S.M.-M.D and J.-C.C. acknowledge financial support from the European Union's Horizon 2020 Research Project and Innovation Program—Graphene Flagship Core3 (N° 881603), from the Fédération Wallonie-Bruxelles through the ARC Grant "DREAMS" (N° 21/26-116) and the EOS project "CONNECT" (N° 40007563), by the Flag-ERA JTC 2019 project entitled "SOGraphMEM" (ANR-19-GRFI-0001-07, R.8012.19), and from the Belgium F.R.S.-FNRS through the research project (N° T.029.22F). Computational resources have been provided by the CISM supercomputing facilities of UCLouvain and the CÉCI consortium funded by F.R.S.-FNRS of Belgium (N° 2.5020.11). These experiments have been performed at the Nanocharacterisation platform in Minatec, Grenoble. We acknowledge Dr. Matthew Bryan for his help in ptychography data treatments.

# Supporting Information

## 1. HAADF imaging of V-doped WSe$_2$

The HAADF image in **Figure S1a** shows the monolayer V-doped WSe$_2$ with small bilayer regions. The sample contains high density of triangular inversion domains consisting of loops of 4|4P mirror twin boundaries as well as point defects at W sites with lower intensity in the HAADF image. The line profile going through one of the typical point defects (**Figure S1d**) shows a low intensity peak that can be attributed to a substitutional atom of lower atomic number at W site.

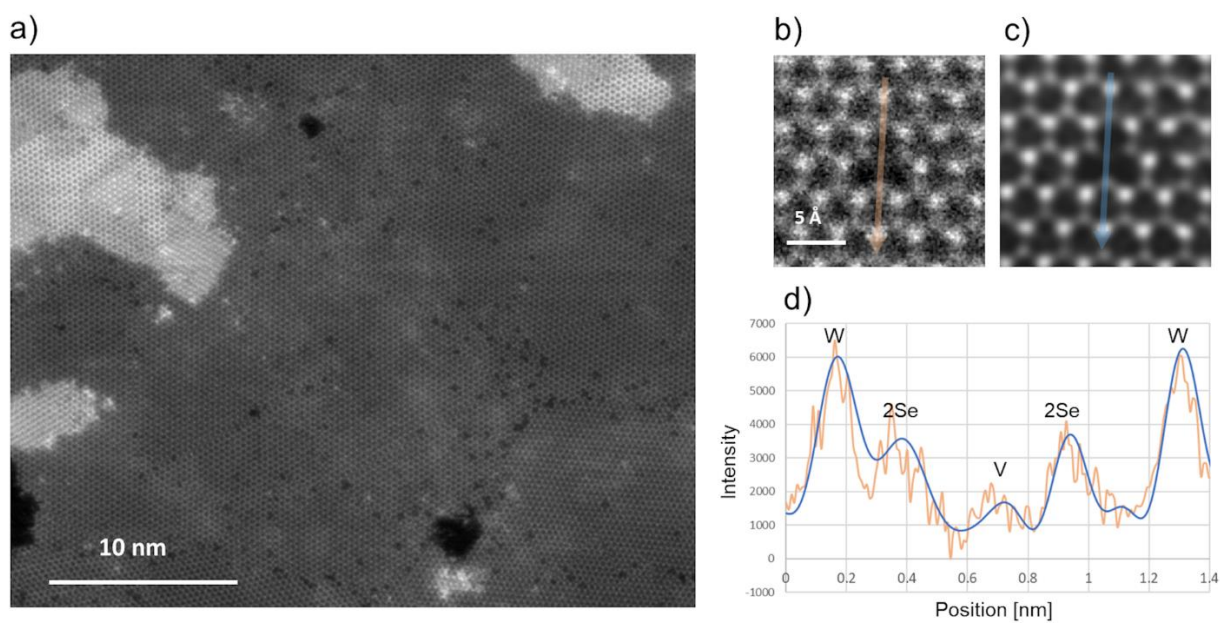

*Figure S1.* (a) HAADF image of V-doped WSe$_2$ showing the layer morphology and typical structural defects. (b) High resolution HAADF image around a typical point defect. (c) Low-pass filtered HAADF image shown in (b). (d) The line profile through HAADF image showing the substitution of W by a lighter element assigned to vanadium.

## 2. Structure of 4|4P mirror twin boundary

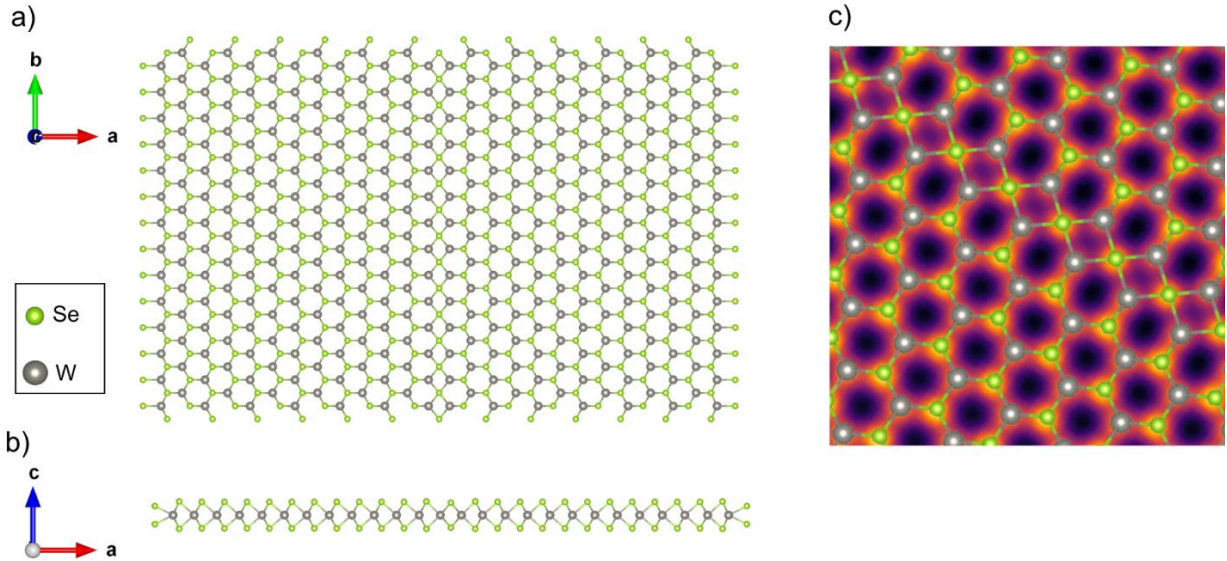

***Figure S2.*** *(a-b): Relaxed structure of a 4|4P mirror twin boundary in WSe$_2$ in plan-view and cross sectional view, respectively. (c) The relaxed structure of 4|4P MTB overlayed on the image of projected potential reconstructed from a simulation with zero lens aberrations.*

## 3. Effect of three-fold astigmatism on DPC-CoM compared to HAADF imaging

The three-fold astigmatism is known to affect the intensities of the atomic columns depending on the relative orientation between the crystal and the electron probe. However, the impact of three-fold astigmatism depends not only on the magnitude and azimuthal angle of the aberration but also on the imaging technique (BF, DF, ADF, HAADF, DPC-CoM). The series of simulations shown in **Figure S3** compares the influence of three-fold astigmatism on the potential maps reconstructed by CoM technique and conventional HAADF images. The same DFT calculated potential of pristine WSe$_2$ is used as input in the three multislice simulations with different aberrations conditions. The first simulation done using a spherically symmetric electron beam shows that contrary to HAADF imaging the intensity of 2Se atomic column is higher than the one of W atoms in potential map reconstructed by CoM. The two following simulations are performed with a constant magnitude of three-fold astigmatism $A_2 = 200$ nm but with two different azimuthal angles of the aberration. The three-fold astigmatism clearly perturbates the intensity of the atomic columns in both potential maps and HAADF images. However, in certain aberration conditions the atomic columns intensity ratio is inversed in the potential maps (**Figure S3h**) where the W atoms appear brighter than the 2Se atomic columns. On the other hand, in the conventional HAADF imaging the W atoms stay brighter than the 2Se atomic columns in all three simulations indicating higher resilience of conventional HAADF imaging to the effect of three-fold astigmatism.

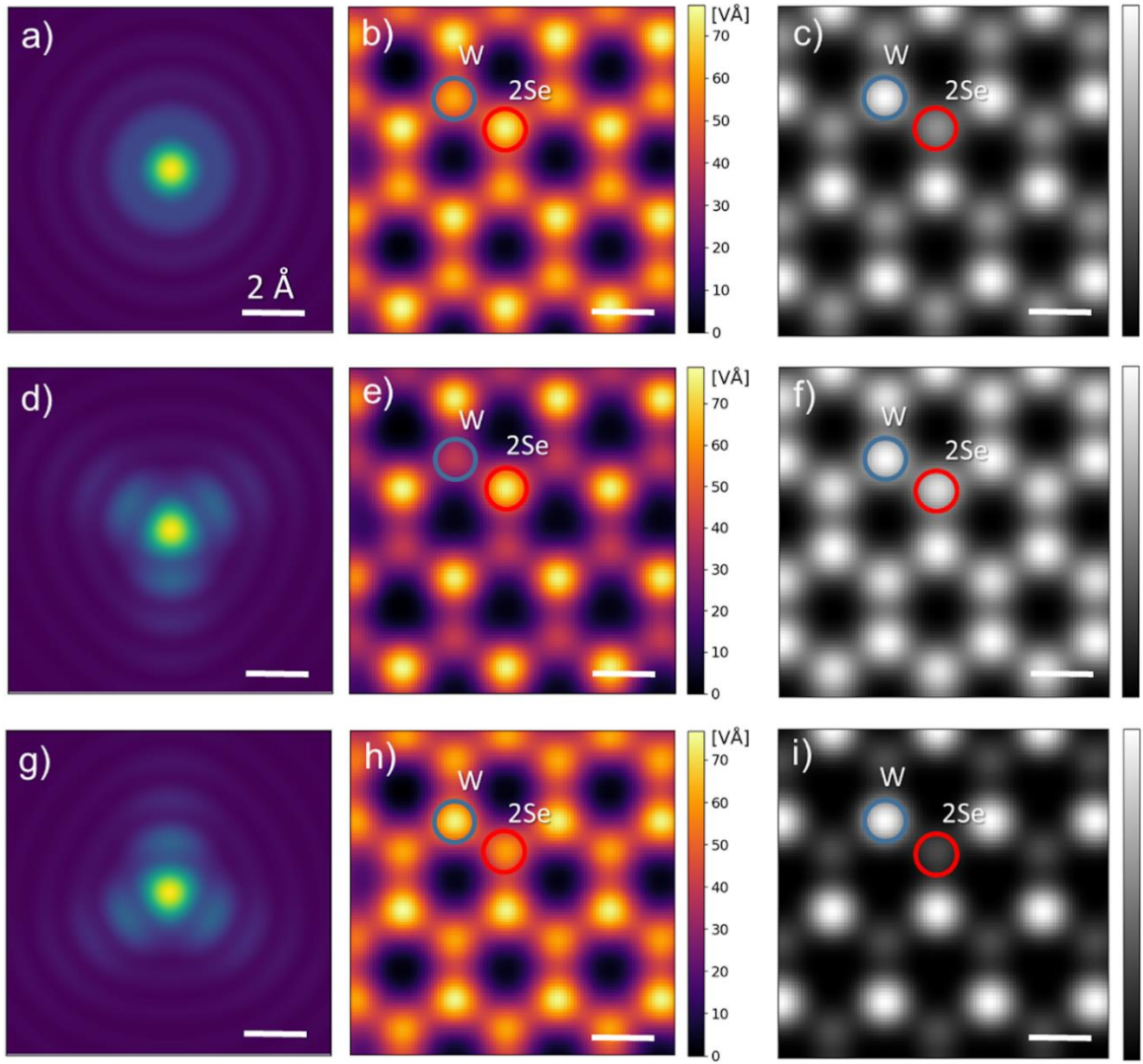

*Figure S3. (a) Image of electron probe simulated with defocus = -9 nm and without any higher order aberrations. (b-c) Corresponding reconstructed potential and HAADF image, respectively. (d) Image of probe simulated with defocus = -9 nm, A2 = 200 nm, $\varphi_{A2}$ = 90°. (e-f): Corresponding reconstructed potential and HAADF image, respectively. (g) Image of probe simulated with defocus = -9 nm, A2 = 200 nm, $\varphi_{A2}$ = 30°. (h-i): Corresponding reconstructed potential and HAADF image, respectively.*

# 4. Dopant charge state visibility in the reconstructed images of projected electric field, potential and charge density

The first physical quantity we extract from CoM images is the vector of projected electric field, from which by means of integration and derivation we can calculate the images of projected potential and projected charge density, respectively. In the following we investigate the visibility of vanadium dopant charge state by comparing two DFT-based multislice STEM simulations: i) without additional charge added to the system i.e. neutral V dopant state shown in **Figure S4a-c** and ii) with a net negative charge of $q = 0.45e$ assigned to the dopant atom, shown in **Figure S4d-f**. The resulting images of projected electric field, potential and charge density reconstructed from the two simulations are compared by calculating difference maps shown in **Figure S4g-i**. The negative charge around the dopant indeed affects the reconstructed electric field, potential and charge density. However, its effect is most easily detectable in the image of projected electrostatic potential where a point charge around single V dopant induces a long-range perturbation centered around the dopant atom with a higher contrast than in the respective images of reconstructed electric field and charge density.

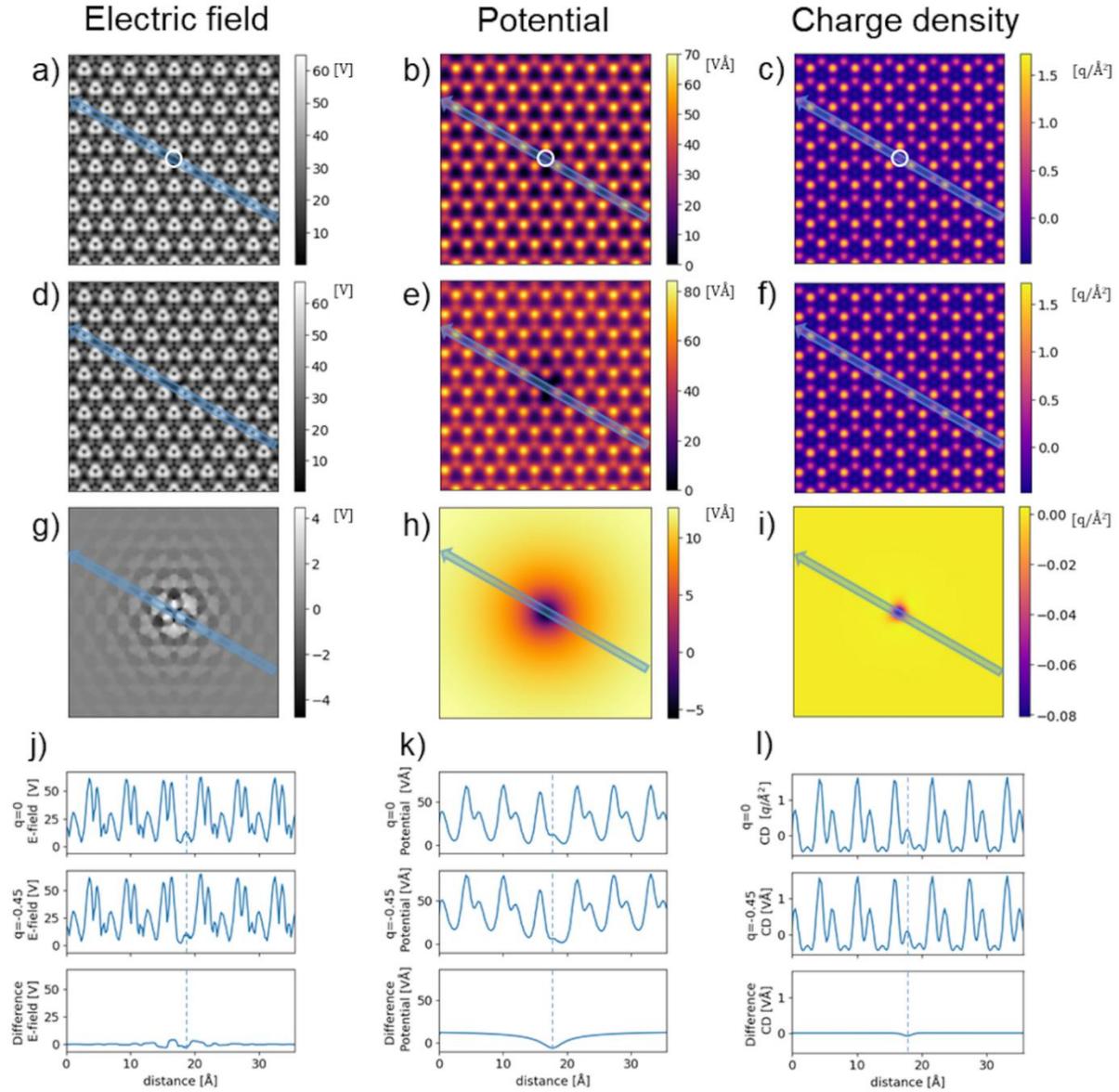

*Figure S4.* *(a-c) Images of projected electric field magnitude, potential and charge density of WSe$_2$ doped with a neutral V dopant. (d-f) Images of projected electric field magnitude, potential and charge density of WSe$_2$ doped with a negatively charged V dopant. (g-i) Difference images of projected electric field magnitude, potential and charge density. j): Line profiles through the regions indicated in a), d) and g) (from top to bottom). k): Line profiles through the regions indicated in b), e) and h) (from top to bottom). l): Line profiles through the regions indicated in c), f) and i) (from top to bottom). The dashed line in line profiles denotes the position of V dopant atom.*

## 5. Neutral vanadium dopant

Although the majority of the observed vanadium dopants exhibit a potential drop assigned to the presence of negative charge, other point defects at W site whose intensity corresponds to vanadium substitutes are occasionally found not to exhibit a substantial variation of electrostatic potential in their vicinity. An example of such point defect assigned to neutral V dopant is shown in **Figure S5a**. Potential map shown in **Figure S5b** is reconstructed from a multislice simulation based on independent atom model where the vanadium dopant is inherently neutral. The line profile comparison (**Figure S5c**) shows that the peak going through the point defect corresponds to the vanadium substitute, however no potential drop converging to this V dopant atom was observed, indicating its neutral charge state.

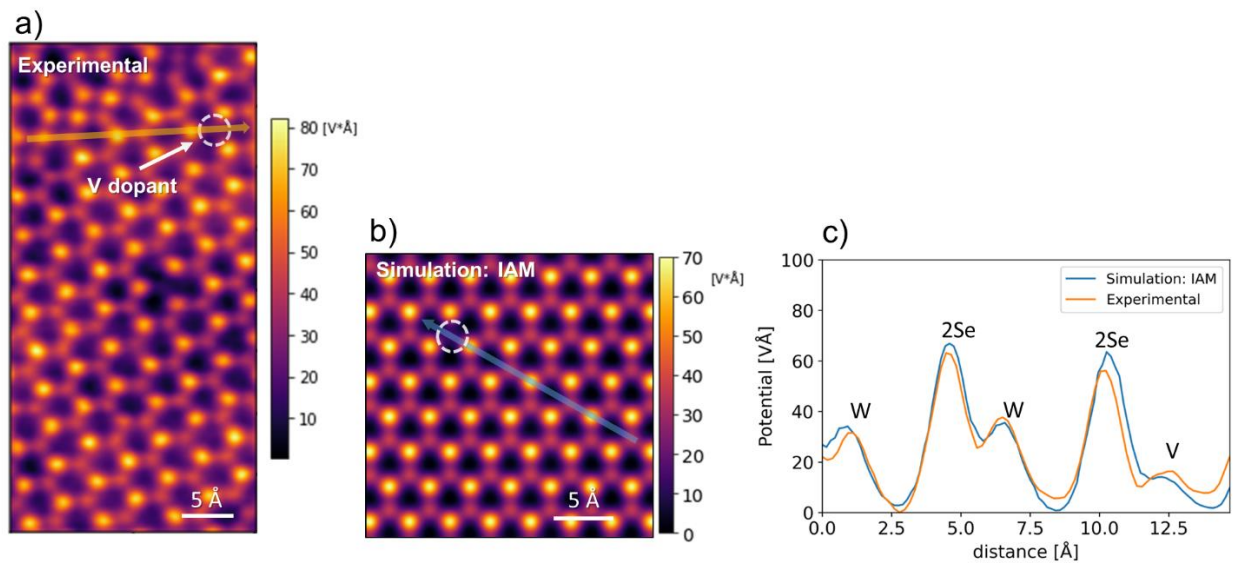

*Figure S5. (a) Experimental potential map with the analysed dopant noted by the white circle. (b) Potential map reconstructed from IAM-based simulation of V-doped $WSe_2$. (c) Comparison of experimental potential line profile going through a neutral V dopant and the potential line profile from IAM-based multislice simulation.*